\documentclass[sigconf,natbib=false,authorversion=true]{acmart}

\author{Vegard Edvardsen}
\affiliation{%
  \institution{Telenor Research}
  \city{Oslo}
  \country{Norway}
}
\email{vegard.edvardsen@telenor.com}

\author{Gard Spreemann}
\affiliation{%
  \institution{Telenor Research}
  \city{Oslo}
  \country{Norway}
}
\email{gard.spreemann@telenor.com}

\author{Jeriek Van den Abeele}
\affiliation{%
  \institution{Telenor Research}
  \city{Oslo}
  \country{Norway}
}
\email{jeriek-van-den.abeele@telenor.com}

\copyrightyear{2022} 
\acmYear{2022} 
\setcopyright{acmlicensed}\acmConference[Q2SWinet '22]{Proceedings of the 18th ACM International Symposium on QoS and Security for Wireless and Mobile Networks}{October 24--28, 2022}{Montreal, QC, Canada}
\acmBooktitle{Proceedings of the 18th ACM International Symposium on QoS and Security for Wireless and Mobile Networks (Q2SWinet '22), October 24--28, 2022, Montreal, QC, Canada}
\acmPrice{15.00}
\acmDOI{10.1145/3551661.3561363}
\acmISBN{978-1-4503-9481-9/22/10}

\usepackage[
  datamodel=acmdatamodel,
  style=acmnumeric,
  sorting=none,
  citestyle=numeric-comp,
  giveninits=true,
  maxnames=1
  ]{biblatex}
\DeclareMathOperator{\loguniform}{LogUniform}
\DeclareMathOperator{\uniform}{Uniform}
\DeclareMathOperator{\relu}{ReLU}

\newcommand\CCSNewLine{
 \texorpdfstring{%
  \global\let\savedtextbullet\textbullet
  \gdef\textbullet{%
    \par\noindent\savedtextbullet\global\let\textbullet\savedtextbullet
  }%
 }%
 {}
}

\title[FORLORN: Comparing Offline Methods and RL for RAN Parameter Optimization]{FORLORN: A Framework for Comparing Offline Methods and Reinforcement Learning for Optimization of RAN Parameters}

\begin{document}

\begin{abstract}
The growing complexity and capacity demands for mobile networks necessitate innovative techniques for optimizing resource usage. Meanwhile, recent breakthroughs have brought Reinforcement Learning (RL) into the domain of continuous control of real-world systems. As a step towards RL-based network control, this paper introduces a new framework for benchmarking the performance of an RL agent in network environments simulated with ns-3. Within this framework, we demonstrate that an RL agent without domain-specific knowledge can learn how to efficiently adjust Radio Access Network (RAN) parameters to match offline optimization in static scenarios, while also adapting on the fly in dynamic scenarios, in order to improve the overall user experience. Our proposed framework may serve as a foundation for further work in developing workflows for designing RL-based RAN control algorithms.
\end{abstract}



\maketitle

\section{Introduction}
\label{sec:introduction}
\begin{figure}
\includegraphics[width=0.8\linewidth]{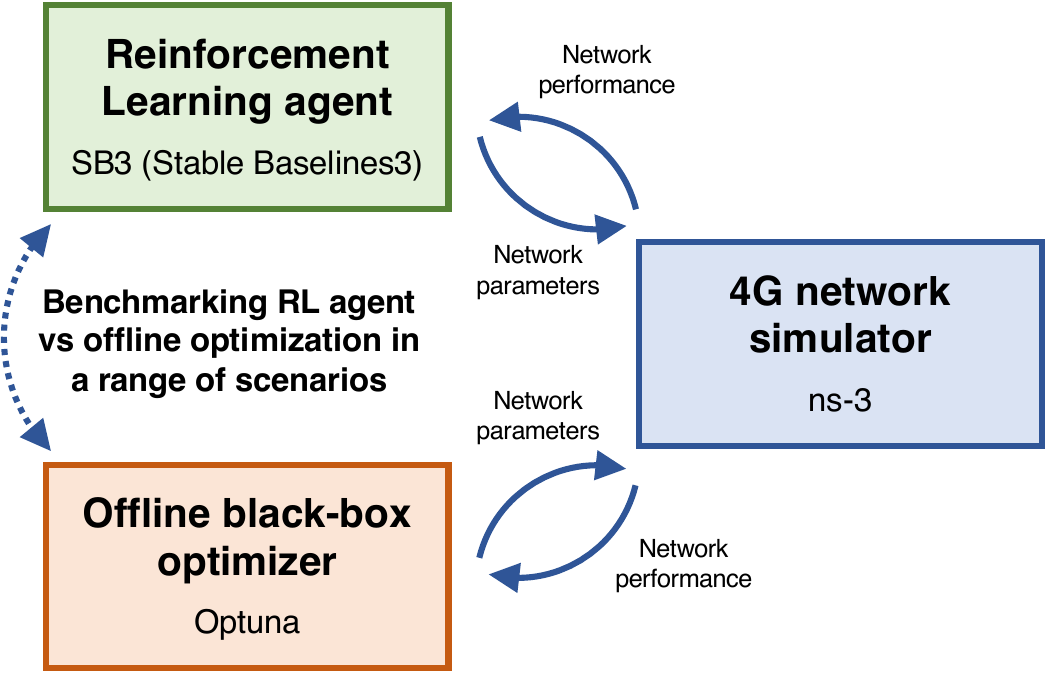}
\caption{Concept of FORLORN: Comparing offline black-box optimization and Reinforcement Learning for RAN parameter optimization in the network simulator ns-3.}
\label{fig:concept}
\end{figure}

A key challenge for mobile network operators lies in coping with the ever-increasing capacity demands on their Radio Access Networks (RAN), as mobile data traffic is expected to continue to grow exponentially. With the introduction of 5G technology, the challenges of maintaining and operating the network escalate further. The network will have to cater to a more heterogeneous set of devices and performance expectations. Furthermore, there will be more cells in the network, and these will have more controllable parameters, such as for configuring new features like beamforming.

Operators seeking to keep costs under control are therefore facing a dual challenge. To save on capital expenses, fine-tuning the base-station parameters in the network could improve the service quality without as much new investment. However, manually fine-tuning these parameters---increasing in numbers, and under continually changing network conditions---would increase headcount and operational expenses. We believe that properly dealing with the complexities of RAN parameter fine-tuning in commercial mobile networks requires an automated optimization approach.

Self-Organizing Networks (SON) functionality has been available from vendors for a while, but now there is a resurgence of interest in the topic, as the emerging Open RAN standards are poised to disaggregate and open up the internals of the base-station stack.
The O-RAN Alliance envisions the RAN Intelligent Controller (RIC) component to play a pivotal role in Open RAN networks~\cite{Polese2022}. RIC will have deep interfaces into base-station internals, and should enable network operators to embed custom automation use-cases in the RAN stack in a standardized fashion. Meanwhile, there is also great interest in adopting automation use-cases based on Machine Learning (ML) methods, and RIC provides an avenue for this.

The paradigm of Deep Reinforcement Learning (DRL) has played a key role in many impressive ML breakthroughs in recent years. Based on training neural networks through trial-and-error learning, DRL has produced state-of-the-art results, such as playing Atari video games~\cite{Mnih2015Feb}, beating humans at board games like Go and chess~\cite{Schrittwieser2020Dec}, and controlling magnetic fields in tokamaks~\cite{Degrave2022Feb}. The adoption of RIC will also enable implementing DRL-based algorithms in Open RAN networks. RIC supports near-real-time closed-loop control down to $10$ milliseconds~\cite{Polese2022}, and provides a rich action space for Reinforcement Learning (RL) agents.

A major hurdle in designing RL-based control algorithms is the training process for the RL agent. As RL is based on trial-and-error learning, the agent needs an \emph{environment} to repeatedly interact with for data acquisition. Having a production network serve as the agent's training environment could impact customers adversely and is thus not viable. Hence, network simulators like ns-3~\cite{Henderson2008} emerge as important tools for developing RL agents, as the simulated environments can provide safe testbeds for the agents to learn by experimenting with various configurations.

However, in the process of developing RL-based algorithms, we need to consider how to benchmark the performance of the RL agent. As the simulator training progresses and the agent's performance converges to a stable level, how should we then assess and validate the performance of the agent?

\begin{figure}
\includegraphics[width=\linewidth]{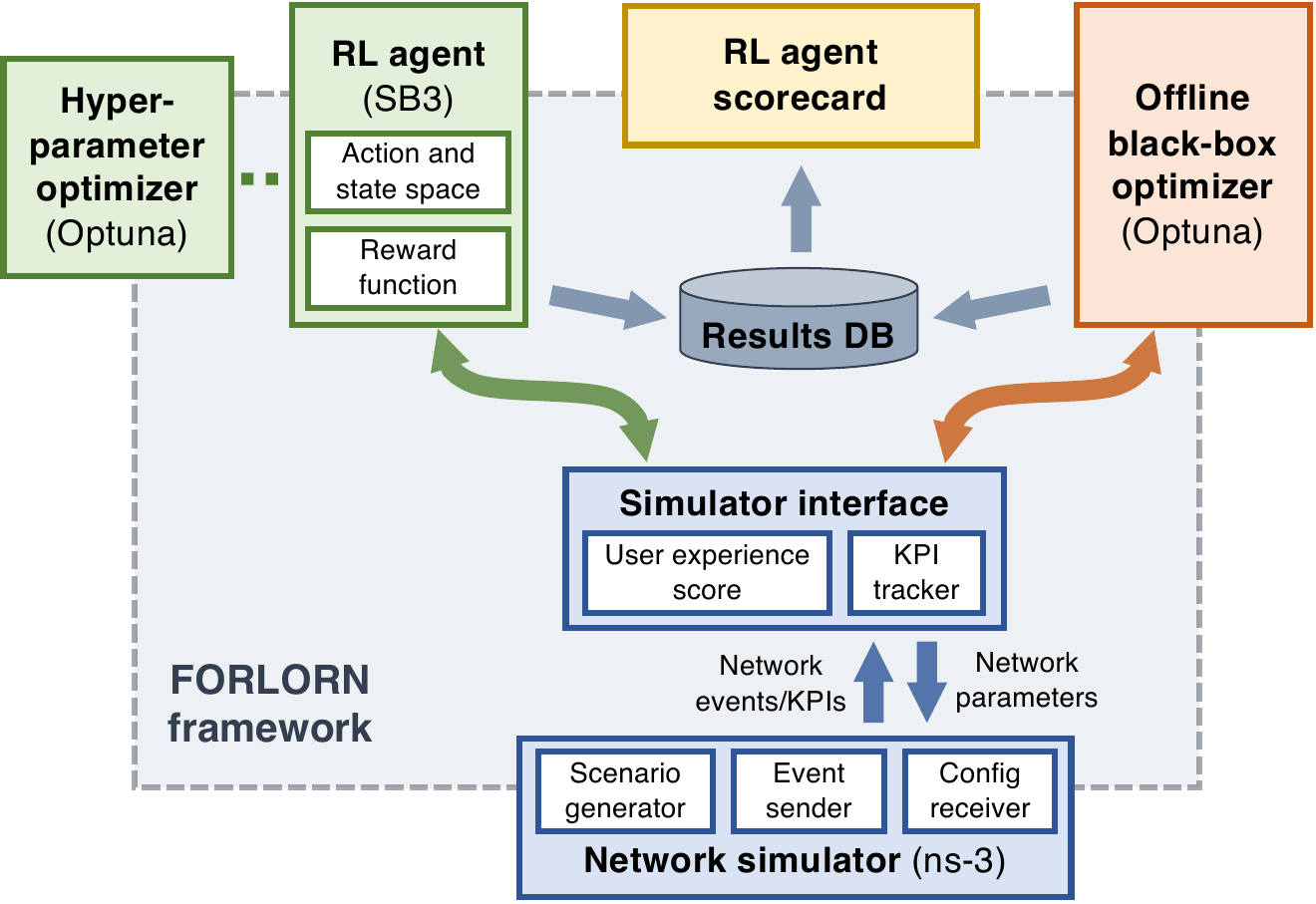}
\caption{Detailed overview of FORLORN, showing how the framework integrates ns-3, SB3 and Optuna to enable a workflow for the development and benchmarking of RL agents.}
\label{fig:overview}
\end{figure}

To establish this benchmark, we propose using \emph{offline black-box optimization} to provide a non-RL baseline. Specifically, we propose using the Optuna~\cite{Akiba2019Jul} framework---popular in the ML community for automated optimization of opaque ``hyperparameter'' settings---to provide the baseline. This RL development workflow is conceptually illustrated in Fig.~\ref{fig:concept}, where the RL agent and the black-box optimizer, both engaging with the same network simulator, are compared against each other.

In this paper, we present FORLORN: a Framework for Comparing Offline Methods and Reinforcement Learning for Optimization of RAN Parameters. The framework is based on integrating the network simulator ns-3 with Stable-Baselines3 (SB3)~\cite{sb3} for training RL agents and Optuna for offline black-box optimization. FORLORN consists of an open-source Python/C++ code base\footnote{Code will be made available at \url{https://github.com/tnresearch/forlorn}.}, and we hope this can spark further work in establishing pipelines for developing and benchmarking RL agents for networking use-cases.

To demonstrate how we envision this framework to be used, we present an example of optimizing the transmission power levels in a 4G network. Essentially a load balancing use-case, such optimization illustrates how AI/ML-based automation in RAN could improve customer experience.


\section{Related Work}
\label{sec:related-work}
There is a rapidly growing research literature exploring the use of DRL in mobile networks, both in network simulators, in wireless testbeds, and in the context of next-generation mobile networks built on Open RAN architecture.

\paragraph{ML in mobile networks}
The application of AI/ML methods to networking use-cases, including DRL as investigated in this paper, has garnered a lot of interest in recent years. Numerous surveys give an outline of this work, e.g., in the context of specific SON use-cases~\cite{Klaine2017}, seen across the specific layers of the networking stack~\cite{Ahmad2020}, or viewed from the perspective of Open RAN architecture~\cite{Brik2022}. Use-cases span the stack, from the lower levels of the physical layer~\cite{Oshea2017}, up through radio resource management use-cases such as traffic scheduling~\cite{Chinchali2018} and handovers~\cite{Lee2020}, to high-level use-cases such as adaptive video streaming~\cite{Mao2017} in the application layer.

For demonstration purposes, in Sec.~\ref{sec:experimental-setup} we present an example use-case of optimizing transmission power among three base stations (``eNBs''). This concept is similar to the work of Alsuhli et al.~\cite{Alsuhli2021IEEE,Alsuhli2021CCNC,Alsuhli2021WCNC}, who investigate RL-based mobility load balancing in ns-3. Their approach is based on adjusting Cell Individual Offsets (CIOs) between neighboring cells, testing several RL methods such as DDQN, DDPG, TD3 and SAC~\cite{Alsuhli2021IEEE}. In extensions to this approach, they also allow the RL agents to adjust cell transmission powers jointly with the CIOs, either in discrete values~\cite{Alsuhli2021CCNC} or continuously~\cite{Alsuhli2021WCNC}.

While the main point of our paper is not the specific use-case, but the approach generally of using an offline black-box optimizer for benchmarking RL agents, we briefly note there are additional differences. For example, while the RL agents of Alsuhli et al. operate with action spaces where settings for all cells can be fully reconfigured in every timestep, our agent (as described in Sec.~\ref{sec:experimental-setup}) is only allowed to make incremental adjustments in each timestep.

\paragraph{Use of network simulators/testbeds for RL}
As the development of RL agents is inextricably linked to the environment in which they are trained, there is much interest in how network simulators and testbeds can be equipped for this purpose. OpenAI Gym is a popular abstraction layer for connecting RL environments to RL algorithms, and ns3-gym~\cite{Gawlowicz2019Nov} provides a toolkit for building such Gym environments in ns-3. Whereas ns3-gym uses ZeroMQ for interprocess communication (IPC), a similar project, ns3-ai~\cite{Yin2020Jun}, proposes instead to use shared memory for high-speed IPC.

For higher-fidelity end-to-end simulations down to the RF level, laboratory testbeds using software-defined radio (SDR) hardware are now starting to be used for RL training. ColO-RAN~\cite{Polese2021} demonstrates an RL agent training in the large-scale wireless network emulator Colosseum. A similar testbed, Powder, is meanwhile being used to investigate RIC use-cases under the NexRAN project~\cite{Johnson2022}.

\paragraph{Hosting RL agents in Open RAN}
RIC in Open RAN is designed to eventually host RAN automation apps such as RL agents~\cite{Niknam2020, Bonati2021,Polese2022}. Accordingly, there is now an interest in how to realize workflows for training and operating RL agents in Open RAN networks. OpenRAN Gym~\cite{Bonati2022} aims to provide a toolbox for developing ML-based RIC use-cases on SDR platforms. Li et al. coin the term ``RLops'' for the management of RL agent life-cycles in Open RAN~\cite{Li2021}.

\section{System Architecture}
\label{sec:system-architecture}
This section presents the problem of RAN parameter optimization, gives an overview of the architecture of the FORLORN framework, and then describes each of the main components in more detail.

\begin{figure}
\includegraphics[width=0.9\linewidth]{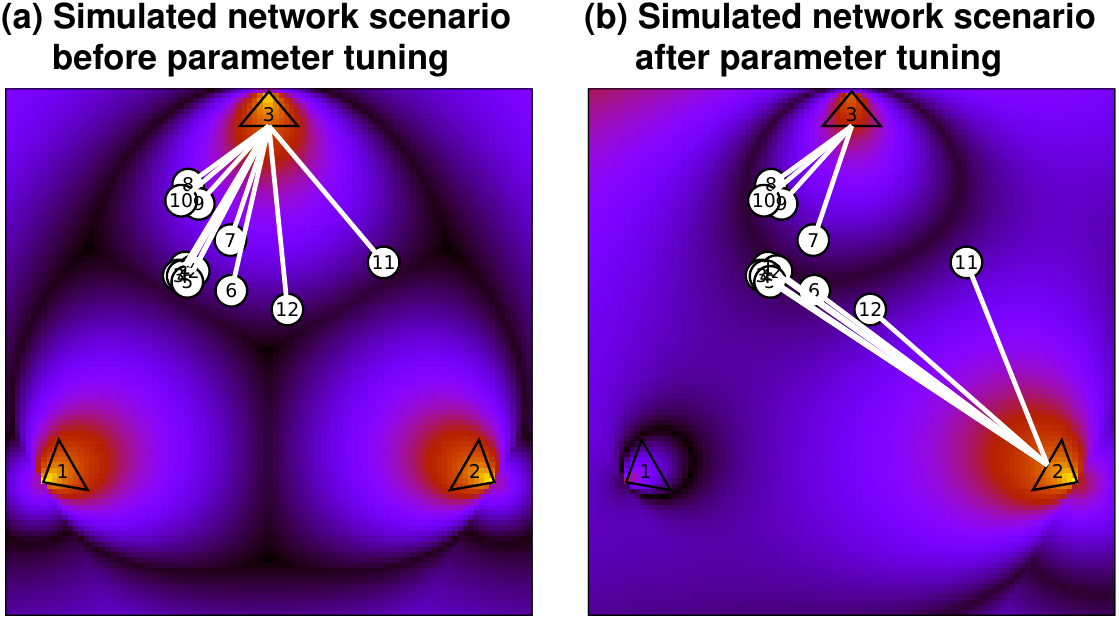}
\caption{Network scenario. (a) Before tuning, all 3 eNBs (triangles) transmit equally strongly. The 12 UEs (circles) are assigned to the same eNB. (b) After tuning, UEs are split among several eNBs. [This figure corresponds to TS2 in Fig.~\ref{fig:optuna_plots}ab.]}
\label{fig:scenario}
\end{figure}

\subsection{Problem: Optimization of RAN parameters}

The problem we consider is how to optimize the parameters of eNBs in a RAN. These parameters are the tunable settings of each eNB, e.g., handover thresholds, power levels and scheduling priorities. Our thesis is that, given the dynamic and heterogeneous nature of network demand, a RAN with continuous fine-tuning of its parameters will more capable of providing good service quality as conditions change, without human intervention. An algorithm that continuously monitors network events and KPIs should be able to automatically make configuration adjustments to improve user experience. This is illustrated by the closed control loop between the network simulator and the optimization agent in Fig.~\ref{fig:concept}.

In this work, we focus on optimization of the transmission powers of the individual eNBs. The transmission power of an eNB is an important parameter, and focusing on just that simplifies the interpretability of the optimization agent at this early stage of development. However, we note that the approach in this paper should in principle be extendable to all types of parameters in the eNBs.

\subsection{Overview of FORLORN}

A structural overview of FORLORN is shown in Fig.~\ref{fig:overview}. It is intended to be a complete framework for designing, training and testing RL agents for the RAN parameter optimization problem. The main components connected by FORLORN are the ns-3 network simulator; SB3 for defining, training and running RL agents; and Optuna for offline black-box optimization.

The key contribution of FORLORN is the coherent integration of these elements, and the setting up of a convenient, replicable workflow for evaluating and visualizing RL agent performance across various network scenarios and for comparing the agent performance against baseline configurations obtained offline.

\subsection{Interfacing with the simulator}

While FORLORN is written in Python, the network scenarios are written in C++ for interfacing with the ns-3 simulation library. In contrast to ns3-gym, which uses ZeroMQ for IPC, we use a simple text-based protocol over standard console pipes (stdin/stdout) for IPC. The simulator is hosted by FORLORN as subprocess instances, emitting salient network events to FORLORN over stdout. In return, FORLORN provides updated network parameters over stdin at fixed simulation-time intervals.

The interface component of FORLORN in charge of hosting the simulator processes, is also responsible for parsing the network events and tracking KPIs over time. The simulator interface calculates a \emph{user experience score}, which is ultimately used as the trial score to compare between RL and offline optimization (see Sec.~\ref{sec:scenario-setup}). Note that while the user experience score is related to the reward function used by the RL agent, it may be different, depending on how the RL agent's reward structure is designed.

\begin{table}\centering\small
\caption{Main network simulation details.}
\label{tab:sim}
\begin{tabular}{ll}
\toprule
Number of nodes & $3$ eNBs, $12$ UEs \\
Distance between eNBs & $1000$ m \\
eNB transmission power & $20$--$40$ dBm \\
eNB antenna pattern & Parabolic, beamwidth $70^\circ$ \\
eNB bandwidth & $5$ MHz \\
Frequency Reuse scheme & Hard reuse ($1/3^\text{rd}$ per eNB) \\
Handovers & A2-A4-RSRQ, ns-3 defaults \\
Data bearer RLC mode & Acknowledged Mode \\
UE downlink traffic & TCP, up to $20$ Mbps CBR \\
UE mobility & At rest, or $14$ m/s when moving \\
Simulator warmup duration & $4$ s \\
Training episode duration & $5$--$20$ s (see \texttt{train\_duration} in Table~\ref{tab:hyperparams}) \\
RL agent interaction interval & $100$ ms \\
\bottomrule
\end{tabular}

\end{table}

\subsection{Offline black-box parameter optimization} \label{sec:offline}

For a given network scenario in the simulator, we establish baseline values for the user experience score by means of offline, black-box optimization over the RAN parameters. We specifically employ Optuna~\cite{Akiba2019Jul}, a modern optimization framework with no domain-specific features relating to RAN parameters, agnostic to the rules of RAN parameter management. In Sec.~\ref{sec:results-rl}, the best-performing parameters discovered by Optuna serve as the benchmark for our RL agent. The problem of RAN parameter optimization is similar to hyperparameter tuning in ML research, where the learning process is affected by numerous parameters of unknown or nearly undetectable impact on the final system performance. Thus, it is of great interest in the ML community to find good solutions in an efficient and automated way, and improve upon ad-hoc trial-and-error.

We employ the \emph{Tree-structured Parzen Estimator} (TPE)~\cite{Bergstra2011} approach implemented in Optuna. This is a sequential parameter optimization algorithm that learns from the trial history (stored by Optuna in an SQLite database). In contrast, (quasi-)random and grid search methods do not learn from previous parameter trials. The results from Optuna's TPE will act as the benchmark for assessing the RL agent's performance.

In practice, each parameter $x$ in the configuration space is assigned a configuration prior over a finite range. This prior distribution is either uniform, log-uniform, or categorical.
For every individual parameter $x$ with a uniform prior, the TPE algorithm fits a truncated Gaussian Mixture Model (GMM) $l(x)$ to the set of the $x$-values of the top-performing parameter combinations in each iteration. A second GMM $g(x)$ is fit to the rest of the $x$-values. For a parameter with log-uniform prior, an exponentiated truncated GMM is used, whereas a categorical prior is merely reweighted to reflect the distributions of the best trials and the remaining ones. This sequentially builds up a probabilistic learning model. 
Candidates for the next parameter value to sample are randomly drawn from $l(x)$, and thus biased towards previously successful choices, while the stochasticity allows for limited exploration of other parameter values. The parameter value that is effectively tried in the next iteration is the one that maximizes the probability ratio $l(x)/g(x)$, which corresponds to maximizing the expected improvement.

Furthermore, Optuna is also highly suitable for problems with combined algorithm selection and hyperparameter optimization. Using TPE, it has been demonstrated to outperform other popular hyperparameter optimization tools in such settings, when it comes to quickly finding good solutions~\cite{Shekhar21}. In Sec.~\ref{sec:rltuning}, we therefore make use of Optuna's TPE implementation to also determine the RL algorithm type and its hyperparameters, in addition to its primary use as the black-box RAN parameter optimizer. 

\subsection{RL for online parameter optimization}
To build and train the RL agents, we use SB3's implementations of the RL algorithms A2C~\cite{a3c} and PPO~\cite{ppo}. While these algorithms are provided ready-for-use by SB3, the specific networking use-case must still be implemented in a way that can be utilized by SB3. Specifically, the use-case designer must implement the \emph{observations} provided to the RL agent, the \emph{actions} available for the agent to use, and specify the \emph{reward} used to guide the agent's behavior. These steps are detailed for the power tuning example in Sec.~\ref{sec:rl-design}.

Once the RL agent has been trained, we can run test trials with the agent in the same network scenarios as for Optuna TPE, and store those results in the same SQLite results database. This enables us to produce an \emph{RL agent scorecard}, showcasing the agent's performance in the context of the benchmarks produced by Optuna.

\subsection{Hyperparameter tuning for the RL agent} \label{sec:rltuning}

The performance of RL agents is typically highly sensitive to hyperparameters chosen during training, as well as to seemingly minor ad-hoc implementation details~\cite{shengyi2022the37implementation}. We therefore employ Optuna in a secondary role, namely as a black-box optimizer over hyperparameters used to train the RL agent. We optimize both properties and hyperparameters of the RL agent itself, such as algorithm type and activation function, and settings for the simulator/environment, such as observations produced and initial conditions. Note, as described in Sec.~\ref{sec:offline}, that this use of Optuna as an optimizer for RL agent and simulator environment hyperparameters is entirely separate from our use of it as an offline mobile network optimizer.

\begin{table*}
  \centering
  \small
  \caption{Hyperparameters optimized for the RL agent. The blank fields are not applicable to the given agent type.}
  \label{tab:hyperparams}
  \begin{tabular}{lllll}
    \toprule
    \textbf{Hyperparameter (env.)}                       & \textbf{Description}                                                   & \multicolumn{2}{l}{\textbf{Configuration prior}}                       & \textbf{Choice} \\
    \midrule
    \texttt{train\_duration}                             & Duration of training episode in milliseconds                           & \multicolumn{2}{l}{$\{5000, 10000, 15000, 20000\}$}                    & $10000$\\
    \texttt{randomize}                                   & Random initial transmit powers, instead of default 30 dBm              & \multicolumn{2}{l}{$\{\text{false}, \text{true}\}$}                    & true\\
    \texttt{history}                                     & Number of timesteps in observations ($T$)                              & \multicolumn{2}{l}{$\{1,2,\dotsc,20\}$}                                & $16$\\
    \texttt{step\_size}                                  & Power increment ($\Delta P$, in 10ths of dBm)                          & \multicolumn{2}{l}{$\{1,2,\dotsc,10\}$}                                & $3$\\
    \texttt{num\_rsrq\_quantiles}                        & Number of RSRQ quantiles in observations ($Q$)                         & \multicolumn{2}{l}{$\{1,2,\dotsc,5\}$}                                 & $1$\\
    \texttt{oob\_means\_gameover}                        & Does power out of bounds mean game over?                               & \multicolumn{2}{l}{$\{\text{false}, \text{true}\}$}                    & true\\
    \texttt{oob\_penalty\_factor}                        & Penalty factor for setting power out of bounds                         & \multicolumn{2}{l}{$\{10^{-4}, 10^{-2}, 1\}$}                          & $1$\\
    \midrule
    \textbf{Hyperparameter (SB3)}                        & \textbf{Description}                                                   & \multicolumn{2}{l}{\textbf{Configuration prior}}                       & \textbf{Choice}\\
    \midrule
    \texttt{ent\_coeff}                                  & Strength of entropy regularization in loss~\cite{williams1991function} & \multicolumn{2}{l}{$\loguniform(10^{-8}, 10^{-1})$}                    & $10^{-3}$\\
    \texttt{gae\_lambda}                                 & Strength of generalized advantage estimator~\cite{gae}                 & \multicolumn{2}{l}{$\{0.8, 0.9, 0.92, 0.95, 0.98, 0.99, 1\}$}          & $0.95$\\
    \texttt{gamma}                                       & Discount factor in future reward estimate                              & \multicolumn{2}{l}{$\{0.9, 0.95, 0.98, 0.99, 0.995, 0.999, 0.9999\}$}  & $0.98$\\
    \texttt{learning\_rate}                              & Learning rate for optimizer                                            & \multicolumn{2}{l}{$\loguniform(10^{-5}, 1)$}                          & $3\cdot10^{-5}$\\
    \texttt{max\_grad\_norm}                             & If norm of gradient is greater than this, scale down                   & \multicolumn{2}{l}{$\{0.3, 0.5, 0.6, 0.7, 0.8, 0.9, 1, 2, 5\}$}        & $1$\\
    \texttt{n\_steps}                                    & Number of steps per parallel environment per update                    & \multicolumn{2}{l}{$\{8, 16, 32, 64, 128, 256, 512, 1024, 2048\}$}     & $256$\\
    \texttt{vf\_coeff}                                   & Strength of value in loss                                              & \multicolumn{2}{l}{$\uniform(0, 1)$}                                   & $0.25$\\
    \texttt{activation\_fn}                              & Neural network activation function                                                    & \multicolumn{2}{l}{$\{\tanh, \relu\}$}                                 & $\relu$\\
    \texttt{net\_arch}                                   & Neural network width ($S_i^{\text{val}}=S_i^{\text{pol}}$)                    & \multicolumn{2}{l}{$\{64, 256\}$}                                      & $256$\\
    \texttt{ortho\_init}                                 & Layerwise orthogonal initial neural network weights?                                  & \multicolumn{2}{l}{$\{\text{false}, \text{true}\}$}                    & false\\
    \texttt{n\_envs}                                     & Number of parallel simulator environments                              & \multicolumn{2}{l}{$\{4, 8, 16\}$}                                     & $16$\\
    \texttt{algo}                                        & RL agent type                                                             & A2C~\cite{a3c}                   & PPO~\cite{ppo}                      & PPO\\
    \cmidrule{3-4}
    \texttt{normalization\_advantage}                    & Normalize advantage across minibatch?                                  & $\{\text{false}, \text{true}\}$  & ---                                 & ---\\
    \texttt{use\_rms\_prop}                              & Use RMSProp instead of Adam as optimizer?                              & $\{\text{false}, \text{true}\}$  & ---                                 & ---\\
    \texttt{clip\_range}                                 & Policy loss clipping range                                             & ---                              & $\{0.1, 0.2, 0.3, 0.4\}$            & $0.2$\\
    \texttt{batch\_size}                                 & Batch size                                                             & ---                              & $\{8, 16, 32, 64, 128, 256, 512\}$  & $128$\\
    \texttt{n\_epochs}                                   & Number of epochs                                                       & ---                              & $\{1, 5, 10, 20\}$                  & $20$\\
    \bottomrule
\end{tabular}

\end{table*}

\section{Experimental Setup}
\label{sec:experimental-setup}
This section presents the optimization use-case of tuning transmission power among three eNBs, used to showcase the functionality of FORLORN, as well as our RL agent for this use-case.

\subsection{Network scenario setup}
\label{sec:scenario-setup}

\begin{figure}
  \centering
  \includegraphics{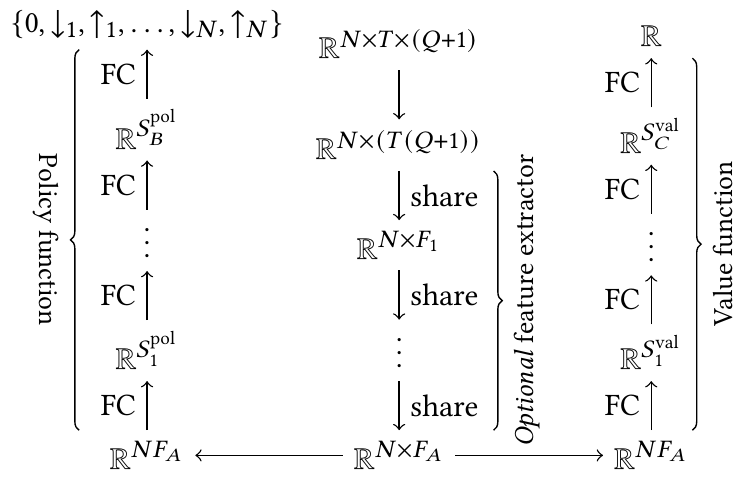}
  \caption{The overall network architecture of our RL agent. $\text{FC}$ denotes a fully connected layer, with all connections having independent weights and biases, while $\text{``share''}$ denotes an FC layer across columns, with weights and biases shared across rows (i.e.\ across eNBs). The feature extraction part is optional and not used in this paper (depth $A=0$). Thus, the raw observations are inputs to both the policy and the value functions, which in that case share no common weights. In our implementation we use the commonly chosen widths $S^{\text{val}}_i=S^{\text{pol}}_i\in\{64, 256\}$ with depths $B=C=2$.}
  \label{fig:network}
\end{figure}

Fig.~\ref{fig:scenario}a shows a top-down view of the optimization scenario. Three eNBs at the vertices of an equilateral triangle point toward the center of the arena. Twelve users (``UEs'') are randomly placed in the arena, with a hierarchical sampling logic that first samples the center, radius and UE count for each UE cluster, and then each UE's location within the clusters. This gives a heterogeneous distribution of UEs across the coverage areas of the three eNBs.

We utilize \emph{hard frequency reuse} in the eNBs, so that the available spectrum is split into thirds for the exclusive use of each eNB. Accordingly, if the transmission power levels are not fine-tuned, user--eNB associations may not be evenly balanced and spectrum will be under-utilized. The aim is then to adjust the transmission power levels, so UEs are load-balanced between the three eNBs (Fig.~\ref{fig:scenario}b). Table~\ref{tab:sim} gives further details on the simulation setup.

\begin{figure*}
\includegraphics[width=0.9\linewidth]{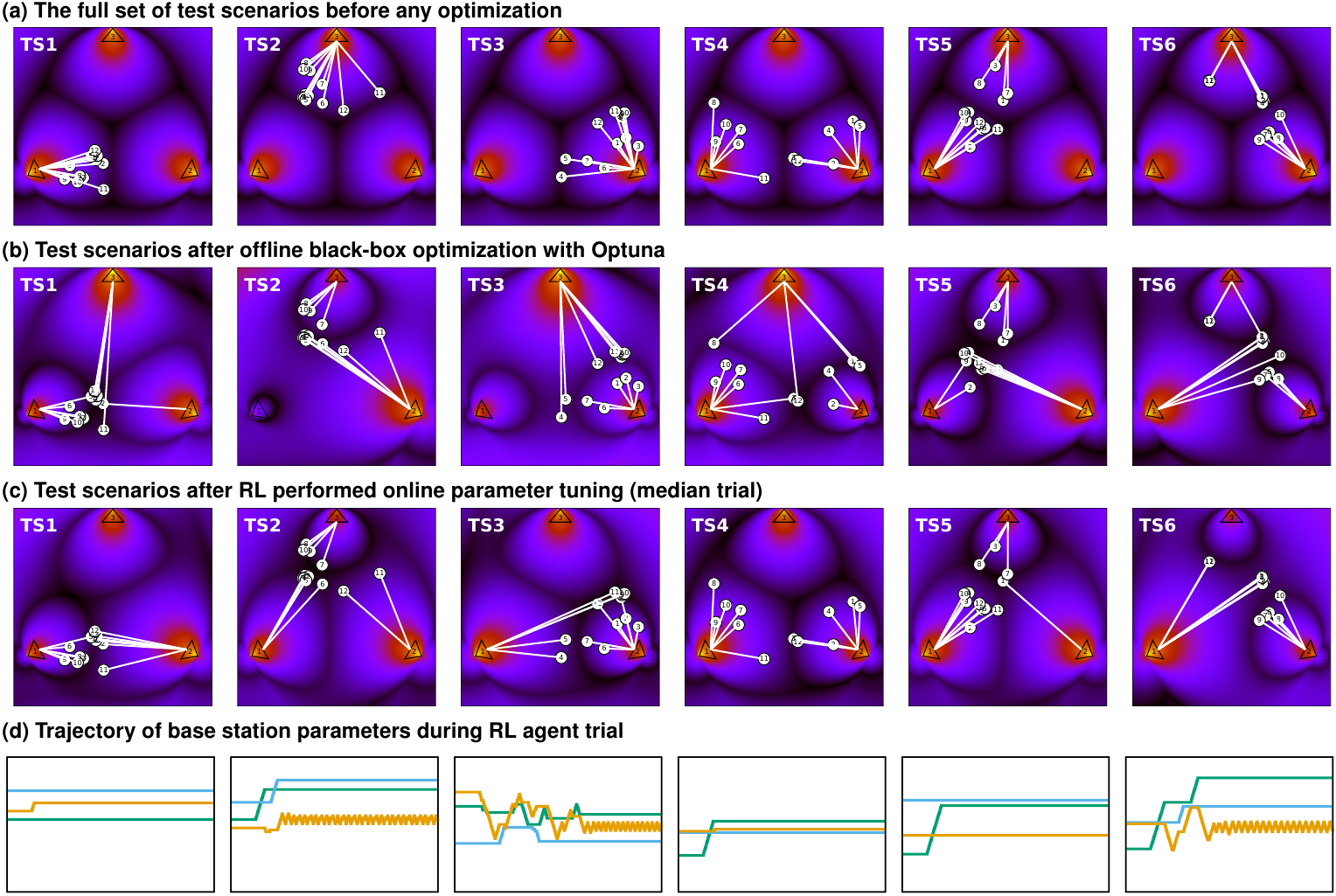}
\caption{Results from example network optimization use-case in the six test scenarios TS1--TS6. (a)~Before optimization. (b)~After 125 trials of Optuna TPE. (c)~Trained RL agent's median-performing trials. (d)~Trajectories of eNB transmission power levels in the median RL trials. [Vertical range 20--40 dBm; green, blue, orange lines for eNB 1--3 respectively.]}
\label{fig:optuna_plots}
\end{figure*}

To benchmark the agent’s performance, we have selected six specific RNG seeds for the UE position sampling algorithm that produces particularly interesting setups; see Fig.~\ref{fig:optuna_plots}a, TS1--TS6, where users are either all located within the coverage area of one particular eNB or two eNBs.

The \emph{evaluation} of a simulation trial is based on a desire for each UE to experience good download throughput, while at the same time penalizing very low throughputs much more than very high ones are rewarded. To this end, we define a custom measure for per-UE experience at a time $t$ as
\begin{equation*}
  q_i(t) = \log_{\alpha}\left(\frac{\left(\alpha-1\right)r_i(t)}{5\cdot 10^5} + 1\right),
\end{equation*}
where $r_i(t)$ is the number of bytes received by UE number $i$ in the two seconds preceding time $t$, and $\alpha=1000$ controls the shape of the function. Thus throughputs above $\sim2 \text{ Mbps}$ ($5\cdot 10^5\text{ bytes}$ in a $2\text{ s}$ window) see diminishing returns, while very small throughputs are heavily penalized in comparison. The trial's total \emph{user experience score} is then the sum of all UE experiences at the final timestep.

\subsection{RL agent design} \label{sec:rl-design}

While having the transmission powers of $N$ eNBs conceptually suggests an agent freely picking actions from the orthant $\mathbb{R}_+^N$, we suspect that highly fluctuating powers are undesirable in practice. Rather than designing a suitably shaped reward that causes the agent to learn to avoid this behavior, we instead choose the discrete action space $\{0, \downarrow_1, \uparrow_1, \dotsc, \downarrow_N, \uparrow_N\}$. For a chosen power increment parameter $\Delta P$, the action with $\downarrow_i$ (resp.\ $\uparrow_i$) then represents turning down (resp.\ up) the transmission power of the $i$'th eNB by $\Delta P$.

The agent reward at a given time is simply the change in the total user experience score since the agent's last environment interaction.
Power settings outside our chosen bounds ($20$--$40$ dBm) are handled by issuing negative rewards and ignoring the action (and optionally terminating the session), not by constraining the action space.

For each eNB, the environment presents the following $Q+1$ observations from the simulator to the agent:
\begin{itemize}
\item Current transmission power ($1$ real number)
\item Number of connected UEs ($1$ natural number)
\item $Q$-quantiles of RSRQs for associated UEs ($Q-1$ real numbers)
\end{itemize}
Keeping a history of observations extending back $T$ timesteps, our observation space becomes $\mathbb{R}^{N\times T \times (Q+1)}$.

The RL agent is structured as a standard actor-critic neural network, and is illustrated in Fig.~\ref{fig:network}. We use the same neural network structure for both the A2C and PPO agents. Both schemes are policy gradient methods wherein the policy distribution function and the value function---used in the advantage function in the loss function---are estimated by separate heads in the same neural network.

Table~\ref{tab:hyperparams} shows the agent and environment (simulator) hyperparameters that we subject to tuning, as discussed in Sec.~\ref{sec:rltuning}. The SB3-specific hyperparameters, along with their priors, are mostly based on the hyperparameter tuning code in RL Baselines3 Zoo~\cite{rl-zoo3}.

\section{Results}
\label{sec:results}
\subsection{Offline optimization}

Fig.~\ref{fig:optuna_plots}a shows the six test scenarios with the default network settings before any optimization has taken place, so that all eNBs are transmitting equally. After 125 trials of Optuna TPE optimization, the best one in each test scenario is as shown in Fig.~\ref{fig:optuna_plots}b.

The final trial scores (i.e. the user experience score) are compared in the scorecard in Fig.~\ref{fig:scorecard}. As discussed in Sec.~\ref{sec:scenario-setup}, when users are clustered at only one or two eNBs, the available spectrum is under-utilized due to hard frequency reuse. Optuna TPE is able to find better configurations that distribute the UEs among more eNBs, resulting in an enhanced overall service quality.

The scorecard in Fig.~\ref{fig:scorecard} also presents a set of \emph{grid search} results, which are the best-performing trials after testing all 125 combinations of the power levels $\{20,25,30,35,40\}$ dBm for each eNB. TPE can surpass grid search with the same number of trials, due to wasting fewer expensive evaluations on trials where important parameters remain fixed. Moreover, TPE adjusts parameters with a finer granularity of $0.1\text{ dBm}$, enabling further improvements.

\begin{figure*}
\includegraphics[width=\linewidth]{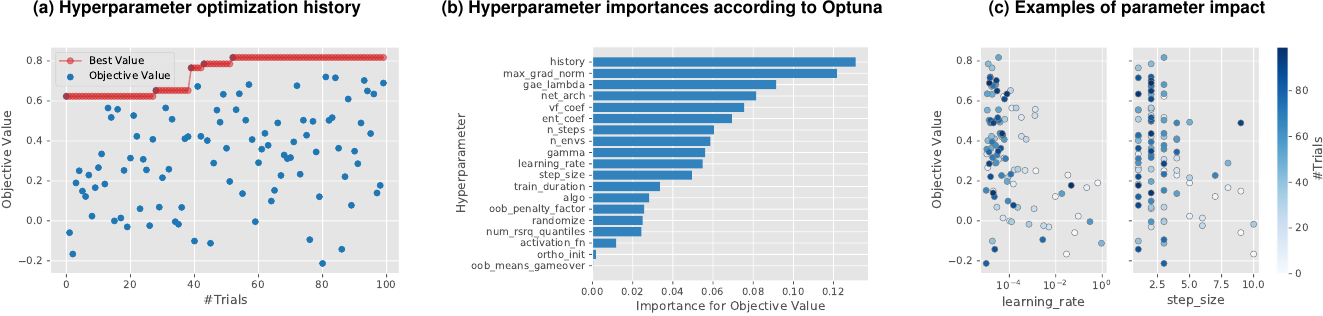}
\caption{Results from RL hyperparameter tuning session. (a)~Model performance across 100 hyperparameter tuning trials. (b)~Indication of the importance of hyperparameters as reported by Optuna, based on fitting regression models to predict trial performance. (c)~Examples of how individual hyperparameters can be inspected to determine impact on model performance.}
\label{fig:hyperparams}
\end{figure*}

\subsection{RL agent performance}
\label{sec:results-rl}

Before training the RL agent to be benchmarked, we first need to fix its hyperparameters.
As described in Sec.~\ref{sec:rltuning}, Fig.~\ref{fig:hyperparams} shows the results from a 100-trial hyperparameter search, each trial training for $2\cdot10^5$ timesteps. Optuna provides a range of visualization capabilities to analyze and understand each hyperparameter's impact on the final performance. Based on these results, we ultimately selected the values listed in Table~\ref{tab:hyperparams}, last column.

The final RL agent was then trained for $10^6$ timesteps. By nature of being an online optimizer, the RL agent may respond differently depending on initial conditions. For each of the test scenarios in Fig.~\ref{fig:optuna_plots}, we therefore ran 100 evaluation trials of the RL agent by setting random initial power levels and letting the agent run for 30 simulation seconds each. The final performance of the agent in these trials is scatterplotted in the scorecard in Fig.~\ref{fig:scorecard}.

As a fair representative of the 100 evaluation trials, we highlight the one with median score in Fig.~\ref{fig:optuna_plots}c. We see that the RL agent selects power settings that more evenly distribute UEs among eNBs, similar to the offline optimizer. This in turn leads to better performance as reflected in the scorecard. Crucially, this is the result of \emph{online, real-time decision-making by the RL agent}, as opposed to Optuna's offline, prearranged optimization process over 125 trials. The RL agent is thus able to continuously adapt to changing network conditions over time. This will be further demonstrated in the next section.

Finally, Fig.~\ref{fig:optuna_plots}d shows the real-time granular adjustments of power levels in the trials from Fig.~\ref{fig:optuna_plots}c. These trajectory plots enable us to see how the agent behaves over time, and we observe rapidly stabilizing relationships between the power levels.

\subsection{Continuous tuning in varying conditions}

A big advantage of the RL-based approach lies in its online nature. Unlike with offline optimization, the agent can dynamically respond to changing conditions. To demonstrate this, we let the UEs move between various positions. Specifically, the users cycle between the positions in TS1, TS2 and TS3, stopping for $30$ sec. in each static configuration. The results are presented in Fig.~\ref{fig:longtrial}.

Fig.~\ref{fig:longtrial}a shows the RL agent's score throughout the long-running test session. The shaded areas correspond to the time periods with the UEs stationary at test scenarios TS1--TS3, showing the interval between the baseline (default settings) and the Optuna TPE best trial. Fig.~\ref{fig:longtrial}b shows the configuration trajectory throughout the session. Note that in between the static test scenarios, there is no corresponding Optuna solution, as the offline optimization would have had to be run on each intermediate setup of user locations.

\section{Discussion and Future Work}
\label{sec:discussion}
The work we have presented in this paper suggests several avenues of further research, regarding improvements to the RL methodology, extensions to the simulated scenarios, and considerations for real-world deployment.

\begin{figure}
\includegraphics[width=0.85\linewidth]{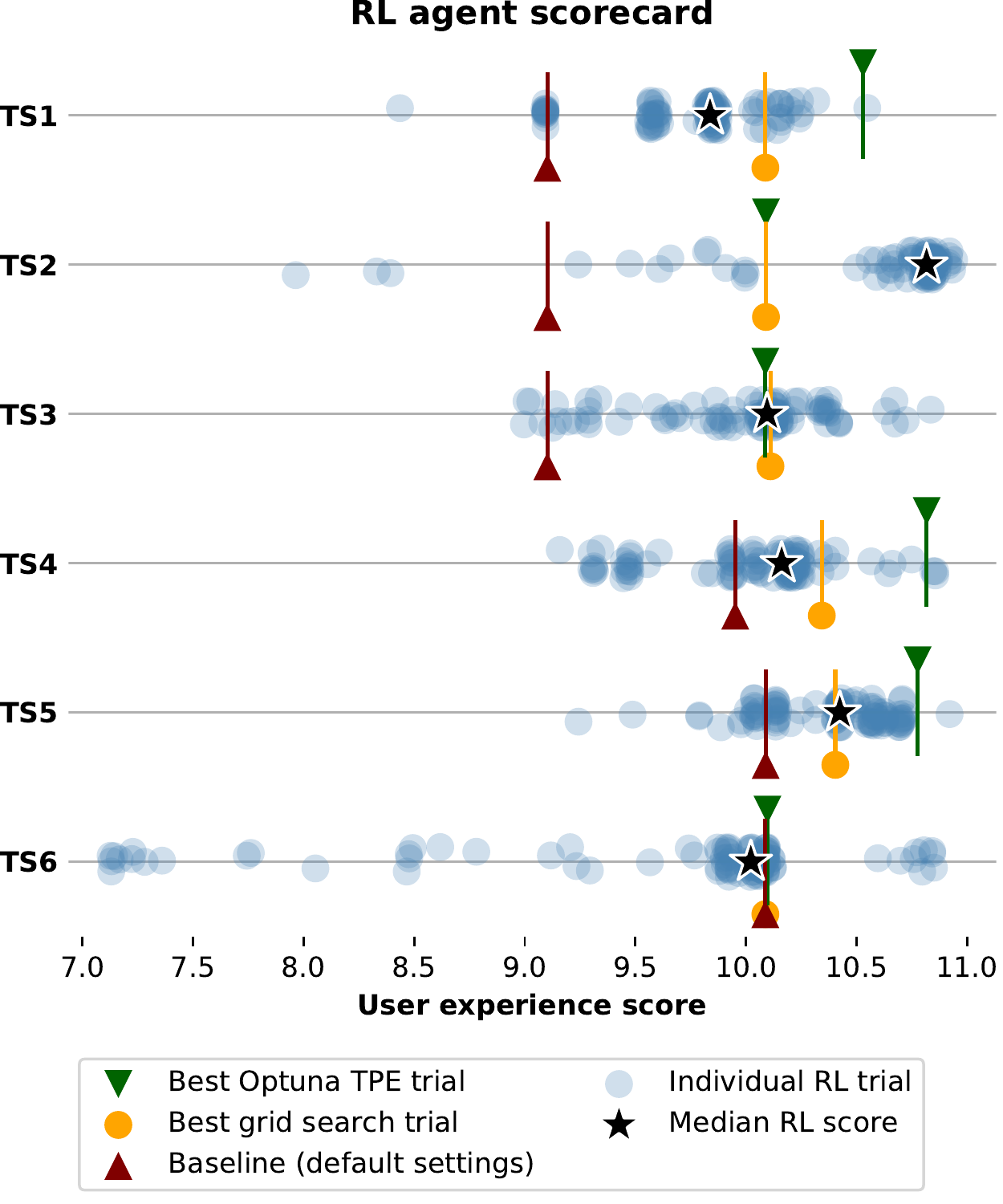}
\caption{The trained RL agent's scorecard, comparing agent performance across 100 test trials in each of the scenarios TS1--TS6 with the offline optimization results. See also Fig.~\ref{fig:optuna_plots}.}
\label{fig:scorecard}
\end{figure}

\paragraph{Richer scenarios and action spaces}
The network optimization scenario in this paper was deliberately kept simple, in order to demonstrate the main concept of using offline optimization methods as a benchmark. However, the real-world potential of such RL agents only comes to fruition when the agent learns to interact with a far richer environment, in terms of higher-fidelity action and observation spaces, user demands, transient traffic patterns, etc. For example, the agent could also be in charge of the frequency reuse scheme itself, handover decisions and scheduling priorities.

\paragraph{RL agent improvements}
For RL agents to address these richer scenarios, further improvements are needed to the agent design itself, such as reward design, feature extraction and neural network architecture. Moreover, having a resilient and interpretable agent is important for real-world deployability. As seen in the RL scorecard in Fig.~\ref{fig:scorecard}, the RL agent currently has some variability in its behavior, which would need to be mitigated in future work.

\paragraph{RL methodology for networking use-cases}
In this work, we have taken the approach of a single RL agent in control of a fixed number of eNBs. It would be impractical to have a single RL agent in control of all eNBs in a geographical area in real settings. At the same time, even our simple transmission power tuning scenario shows the importance of inter-eNB ``collaboration''. To this end, we envision multiple agents in charge of a variable number of eNBs, communicating among themselves, e.g., across the X2 interface. This leads to many questions for future work, including feature extraction for variable numbers of eNBs, shared parameters between eNBs, and perhaps federated learning in order to encompass more state-space exploration than any single eNB is capable of.

\paragraph{Real-world adoption}
Real-world deployment of RL agents for RAN parameter optimization will become possible with the introduction of RAN Intelligent Controller in Open RAN. However, large-scale adoption of RL methods in RAN will be complicated by regulations, training complexities, validation requirements, etc. While much can be investigated in simulators and SDR testbeds, it remains an open question whether the \emph{simulation-to-reality gap} hinders direct deployment to real networks.

\section{Conclusion} \label{sec:conclusion}

In this paper we have presented a framework for comparing RL and offline optimization methods for RAN parameter optimization. To demonstrate the concept, we presented results from a transmission power optimization problem. Using the Optuna-derived benchmark baselines in various test scenarios, we compiled a scorecard to assess the RL agent's performance.

Our results demonstrate that RL agents can compete with offline-optimized results, showing promise that online continuous control of base station parameters can be a viable approach, possibly even in real-world mobile networks with the adoption of Open RAN with RIC. Although we only demonstrate this approach with a few parameters, recent real-world applications of RL illustrate the potential for extensions to more complex action spaces.

As demonstrated in this paper, coupling a network simulator with an offline optimizer for performance benchmarking is a useful strategy for developing RL-based RAN controllers for next-generation mobile networks. The framework presented here can serve as the foundation for future work in this direction.

\begin{figure}
\includegraphics[width=\linewidth]{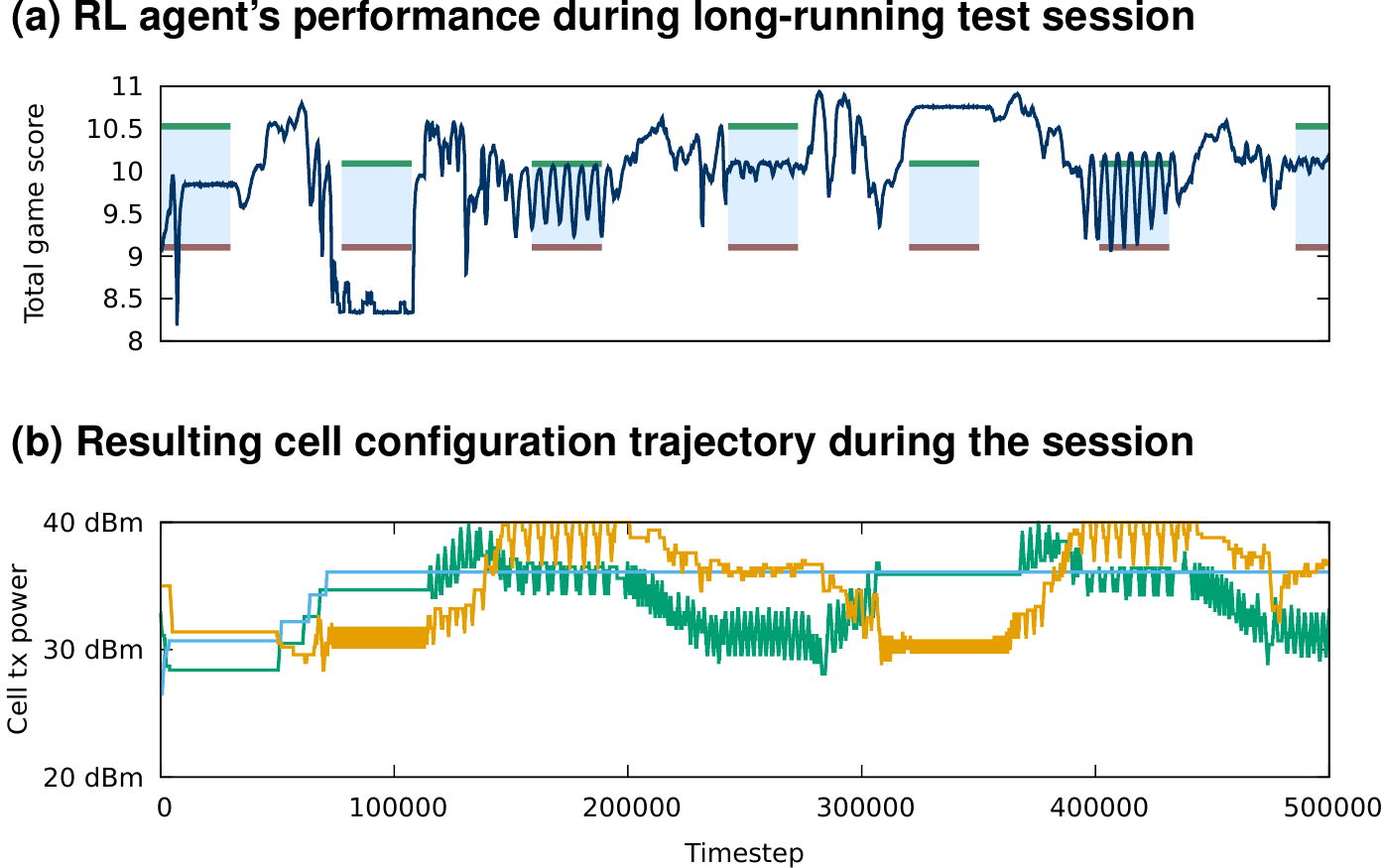}
\caption{RL results from a long-running test trial where users repeatedly cycle through scenarios TS1--TS3. (a)~User experience score over time, superimposed on benchmark scores from Fig.~\ref{fig:scorecard} whenever users are standing still at TS1/TS2/TS3. (b)~Trajectory of eNB power levels throughout the session.}
\label{fig:longtrial}
\end{figure}

\balance
\printbibliography

\end{document}